\documentstyle[graphicx,prb,aps]{revtex}
\def\f{$f$}
\begin{document}
\date{\today}
\title{
A novel method of self interaction corrections in density functional 
calculations}

\author{Urban Lundin and Olle Eriksson} 

\address{Department of Physics, 
University of Uppsala
Box 530, S-751 21 Uppsala, Sweden }

\maketitle

\begin{abstract}
It is demonstrated that the commonly applied self interaction correction
(SIC) used in 
density functional theory does not remove all self interaction. 
We present as an alternative a novel method which, by construction, is
totally free from self interaction. The method has the correct 
asymptotic 1/r dependence. 
We apply the new theory to localized $f$-electrons in
praseodymium and compare with the old version of SIC, the local 
density approximation (LDA) and with an atomic Hartree-Fock calculation. 
The results show a lowering of the $f$ level, a contraction of
the $f$ electron cloud and a lowering of the total energy by 13 eV per 4\f 
electron compared to LDA. The equilibrium volume of the new SIC method
is close to the ones given by LDA and the older SIC method, and is in
good agreement with experiment. The experimental cohesive energy is
in better agreement using the new SIC method, both compared 
to LDA and another SIC method.  
\end{abstract}

\section*{Introduction}
{\bf Published as: Int.\ J.\ Quantum Chem. {\bf 81}, 247 (2001).}\\
Density functional theory (DFT) states that the ground state energy of a many 
particle system is an exact functional of the total particle density 
$\rho$\cite{hohenberg}.
Unfortunately the exact functional is unknown and in order to have a
practical mean of calculating the ground state energy one maps the
interacting particle system onto a non-interacting, for which
the total energy can be written as\cite{kohn}
\begin{eqnarray}
\label{tote}
E[\rho]=T[\rho]+E_{{\rm H}}[\rho]+E_{{\rm XC}}[\rho]+ E_{{\rm NN}}+E_{{\rm eN}}[\rho],
\end{eqnarray}
where $T[\rho]$ is the one-electron kinetic energy 
(given by $\sum n_i \epsilon_i - \int V_{{\rm eff}}\rho$),
$E_{{\rm H}}[\rho]$ is 
the Hartree interaction. $E_{{\rm XC}}[\rho]$, the exchange-correlation (XC) 
potential, is supposed to correct for the approximations 
introduced when replacing the true many-body kinetic energy by 
$T[\rho]$ and the correct electron-electron interaction by 
$E_{{\rm H}}[\rho]$. In addition $E_{{\rm NN}}$ is
the electrostatic nucleus-nucleus interaction, and $E_{{\rm eN}}[\rho]$ is the 
electron-nucleus interaction. 
Since the exact XC potential 
is not known, one has to rely to approximations. The simplest (and most 
frequently used) approximation is the LDA, which is a parameterized
XC functional using
a homogeneous electron density. Approximate expressions for the total
energy combined with efficient methods for solving the Kohn-Sham
equations\cite{kohn} have resulted in powerful methods for studying many
materials properties such as magnetism, chemical bonding, hardness,
elasticity as well as surface related phenomena, including relaxation,
reconstruction and chemisorption. Quite frequently LDA
calculations reproduce experimental results with an error smaller than a
few percent. However, notable exceptions exist.
One problem is that the approximations leading to the LDA 
introduce an unphysical
self interaction (SI) of a particle with itself. 
For instance, in the local density approximation the Hartree part
contains the mean-field interaction energy of an electron with itself.
An exact XC functional
would have a SI part which would exactly cancel 
the SI of the Hartree term. The result
would be a theory totally free from 
self interaction, giving the exact ground state energy.
It may be noted here that for systems with broad bands, where the electron
states form Bloch states, self-interaction becomes neglibly small, whereas
for systems where the electrons are localized or close to localization the 
self-interaction problem becomes serious.
A self interaction correction (SIC) method of DFT was
presented in 1981 by Perdew and Zunger\cite{perdew} (here referred to as PZ). 
Although there exist improved schemes\cite{vosko,cortona} the PZ-method is 
the method predominantly used. One of the virtues of the
correction of PZ is that for the exact XC functional
the added correction becomes zero.
The method of PZ has been applied with some success to different systems,
e.g. localized electrons in Ce and
Pr\cite{beiden,svane1,szotek1,svane2,svane3,temmerman,szotek2,forstreuter},
a 1-D Hubbard
model\cite{1-dhubb} and transition metal mono oxides\cite{jones,oxides,arai}.
Although the PZ version of correcting for SI has been successful in
many applications we question its ability to remove all self interaction
and suggest, as an alternative, a novel way of performing LDA-SIC
calculations, which is totally free from self interaction.

\section*{Exchange and correlation in self interaction corrections}
The self interaction corrected XC 
energy, within the approach by Perdew and Zunger\cite{perdew}, 
applied to LDA, is normally written as 
\begin{equation}
\label{wrong}
E_{{\rm XC}}^{{\rm SIC}} [n]=
E_{{\rm XC}}^{{\rm LSD}} [n]-\sum_{i\sigma}^{{\rm occ}} \left[ \frac{1}{2}
\int d^3 r d^3 r' \frac{\rho_i^\sigma ({\bf r})\rho_i^\sigma ({\bf r}')}
{\mid {\bf r - r}' \mid}   +  \frac{1}{2}
\int d^3 r \epsilon_{{\rm XC}}(\rho_i^\sigma ({\bf r}),0)
 \rho_i^\sigma ({\bf r})\right].
\end{equation} 
The basic idea behind this equation is to subtract the interaction energy
of an electron with itself from the total energy. The second 
term on the right hand side of
Eqn.(\ref{wrong}) corrects for the self-Hartree interaction but we choose to
include this correction to the XC functional
since formally one may introduce all terms there which are not found in the
kinetic energy functional and the Hartree functional.  
The last term on the right hand side
of Eqn.(\ref{wrong}) is called the self XC. The 0 in
$\epsilon_{{\rm XC}}(\rho_i^\sigma ({\bf r}),0)$ indicates that 
the correction is made separately for each spin-channel.  
Minimization of the SI corrected energy functional with respect to the density
leads to Kohn-Sham like equations with a SI corrected effective
potential, $V_{{\rm eff}}$. The corrections coming from SI are normally
referred to as the SIC part of the potential.
The correction for the
self Hartree interaction is exact, but as will be argued below the
correction for self XC is imperfect.
To illuminate the problem, we
consider an LDA functional (Eqn.(\ref{tote}))
and take a closer look at the XC functional.
It is written as
$E_{{\rm XC}}(\rho({\bf r}),0)=\int d^3r \epsilon_{{\rm XC}}(\rho({\bf r}),0)
\rho({\bf r})$, 
where one term in the XC energy density,
$\epsilon_{{\rm XC}}$, 
is proportional to $\rho^{1/3}$. Let us now write the total density as
$\rho_{{\rm tot}}=
\bar{\rho}+\rho_i^\sigma $, where
$\rho_i$ is the density given by the orbital which we are interested to
correct for self interaction, $\bar{\rho}$ is
the summed density of all other orbitals and $\rho_{{\rm tot}}$  the total 
electron density. It is now easy to see that 
the correction of PZ, as defined in Eqn.(\ref{wrong}), does 
not remove all self XC, since with the 
above expressions for $\rho_{{\rm tot}}, \bar{\rho}$ and
$\rho_i^\sigma $, we obtain from Eqn.(\ref{wrong}) an expression for the 
exchange part of the energy functional
\begin{equation}
\label{rho3}
\int d^3r (\bar{\rho}+\rho_i^\sigma )^{4/3} - 
\int d^3r(\rho_i^\sigma)^{4/3}.
\end{equation}
Eqn.(\ref{rho3}) is not equal to $\int d^3r (\bar{\rho})^{4/3}$ and hence does 
not represent a contribution to the energy functional that is free 
from self interaction. 

In a similar way the correction of PZ does not remove the self XC
interaction in the effective potential of the Kohn-Sham equations.
The XC part of the
effective potential, $V_{XC}$, of the Kohn-Sham equations
is in LDA written as\cite{barth}
\begin{eqnarray}
\label{xclda}
V_{{\rm XC}}^{{\rm LSDA}}(\rho_i,{\bf r}_s,x)=\left[\mu_x^P({\bf r}_s)+v_c({\bf r}_s) 
\right] 2\rho_i^{1/3} +
\mu_c^P({\bf r}_s)-v_c({\bf r}_s), 
\end{eqnarray}
where it may be seen that the exchange potential ($\mu_x^P({\bf r}_s)2\rho_i^{1/3}$ 
contains a term proportional to $\rho({\bf r})^{1/3}$ 
(for a description of terms in Eqn(\ref{xclda}),see Ref.~\cite{barth}). 
The PZ corrected effective
potential for electron state $i$ involves, analogous to Eqn.(\ref{wrong}), the following terms,
\begin{equation}
\label{effpot}
V_{{\rm eff},i}^{{\rm PZ}} = 
\int \frac{\bar \rho({\bf r}') } {\mid{\bf r-r}'\mid} d^3 r' +
{\rm A}({\bf r})([{\bar \rho({\bf r})}+ \rho_i^\sigma ({\bf r})]^{1/3}-
{\rho_i^\sigma ({\bf r}})^{1/3}), 
\end{equation}
and we observe that due to the non-linear dependence on the
electron density, $V_{{\rm eff},i}^{{\rm PZ}}$ is dependent on $\rho_i$ and hence 
SI effects are present in the effective
potentials well. This was also pointed out 
by other authors\cite{cortona,nesbet}. 

\section*{A novel method of self interaction}
We have now pointed out why the SIC form of PZ does not remove all the
SI and below we suggest a method which does this.
In general, the approach is to identify all contributions in the LDA energy
expression (Eqn.(\ref{tote})) and the effective potential which contain 
SI and to remove these interactions. 
Writing $\rho_{{\rm tot}}=
\bar{\rho}+\rho_i^\sigma $, one may obtain an energy expression 
which is free from
self-interaction, by applying the correction to the densities instead of 
the energies. The energies for $\rho_i^\sigma $ are calculated from 
$\bar{\rho}$ instead of the total density $\rho_{{\rm tot}}$. 
LDA contains a term $E[\bar{\rho}+\rho_i^\sigma ](\bar{\rho}+\rho_i^\sigma )$ 
whereas the wanted expression is 
$E[\bar{\rho}+\rho_i^\sigma ]\bar{\rho}+E(\bar{\rho})\rho_i^\sigma $ 
($E$ denoting the total energy functional) if we only apply SIC to orbital 
$\rho_i^\sigma$.
The expression for the total energy, when formulated as $E^{{\rm tot}}=E^{{\rm LDA}}+
E^{{\rm corr.}}$, is
\begin{eqnarray}
\label{tote:correct}
E^{{\rm tot}}_{{\rm SIC}}=E^{{\rm tot}}_{{\rm LDA}}(\epsilon^{{\rm SIC}}) -
\frac{1}{2}\sum_{i\sigma} \int  \frac{\rho_i^\sigma ({\bf r})\rho_i^\sigma 
({\bf r}')}
{\mid {\bf r - r}' \mid}d^3 r d^3 r' - \nonumber \\ \sum_{i\sigma}
\int [ \epsilon_{{\rm XC}}(\rho)-\epsilon_{{\rm XC}}({\bar{\rho}}
) ] \rho_i^\sigma   d^3 r + 
\sum_{i\sigma} \int [ V_{{\rm eff}}(\rho)-V_{{\rm eff}}({\bar{\rho}}) ] \rho_i^\sigma d^3 r,
\end{eqnarray}
where $E^{{\rm tot}}_{{\rm LDA}}(\epsilon^{{\rm SIC}})$ is the normal LDA total energy
expression (Eqn.(\ref{tote})) evaluated with a KS-eigenvalue obtained from 
Eqn.(\ref{effpot_new}) below, 
the effective potential.  To arrive at the expression in Eqn.(\ref{tote:correct}) 
we have explicitly subtracted the parts in the LDA expression 
which contain SI and thereafter added a SI free part.

The effective potential to be used in the KS-equation is constructed in the
usual way 
\begin{equation}
V_{{\rm eff}}(\rho,\rho_i^\sigma ,{\bf r})=\frac{\delta [E_{{\rm H}}+E_{{\rm XC}}]}{\delta 
\rho_i^\sigma ({\bf r})}+V_{{\rm eN}},
\end{equation}
where $V_{{\rm eN}}$ is the electron-nucleus potential 2Z/r. 
Explicitly taking the functional derivative, we get the potential
\begin{eqnarray}
\label{effpot_new}
V_{{\rm eff}}(\rho_{tot},\rho_i,{\bf r})=\int \frac{\rho_{tot}({\bf r}')-
\rho_i^\sigma ({\bf r}')}
{\mid{\bf r-r}'\mid} d^3 {\bf r}' + 
{\rm A}({\bf r})[\rho_{{\rm tot}}({\bf r}')-\rho_i^\sigma 
({\bf r}')]^{1/3}+V_{{\rm eN}}= \nonumber \\
\int \frac{\bar{\rho}({\bf r}')}{\mid{\bf r-r}'\mid} d^3 {\bf r}' -
{\rm A}({\bf r}){\bar \rho}({\bf r})^{1/3}+V_{{\rm eN}}.
\end{eqnarray}
This effective potential is by construction totally free from SI if one
uses the electron density ${\bar \rho}$ in the functional A$({\bf r})$. 
The Hartree part in this equation coincides exactly with the 
Hartree part in Eqn.(\ref{wrong}).
In short, the potential for the orbital $\rho_i^\sigma $ can be written as
$V(\rho_{{\rm tot}}-\rho_i^\sigma ,{\bf r})=V({\bar{\rho}},{\bf r})$,
instead of the normal orbital independent potential 
$V(\rho_{{\rm tot}},{\bf r})$, or the PZ-SIC: 
$V(\rho_{{\rm tot}},{\bf r})-V(\rho_i^\sigma ,{\bf r})$.
This procedure is applicable to {\it all} approximations for the
XC energy but here, for simplicity, we choose to discuss the LDA. 
In the asymptotic limit ${\bf r} \rightarrow \infty$ the effective 
potential has the correct $1\over r$ dependence, since we subtract one 
electron from the screening cloud of the neutral atom. One can arrive at 
Eqn.(\ref{effpot_new}) by starting from the Hartree-Fock
expression for the effective potential, which is free from self interaction, 
and then replace the Fock part by
A$({\bf r}){\bar \rho}({\bf r})^{1/3}$. This replacement then hopefully 
both simplifies the
effective potential by making it local as well as includes correlation effects.

\section*{Numerical tests on the method}
\begin{figure}[hbt]
\includegraphics[width=0.9\textwidth]{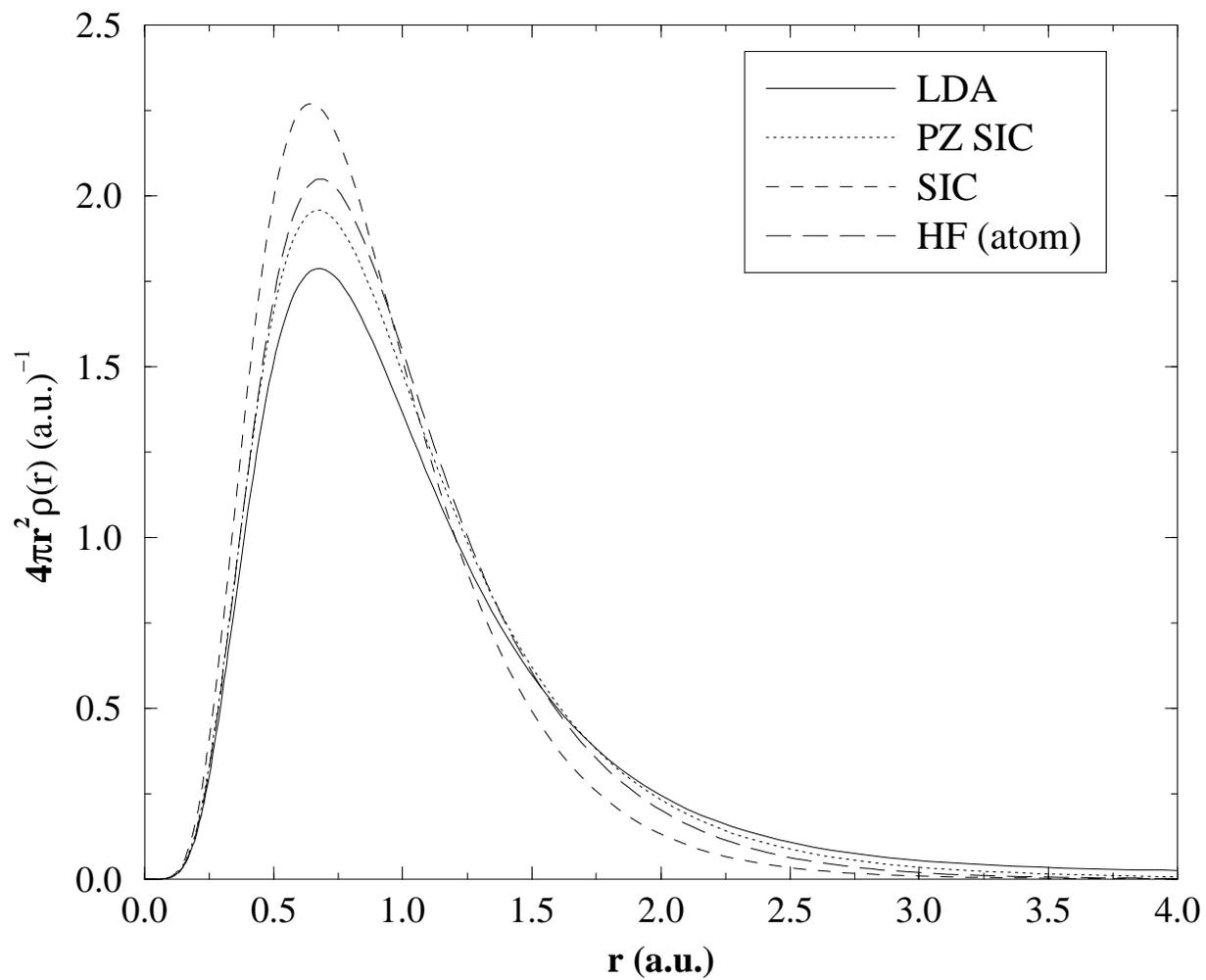}
\caption{The 4$f$ radial density using different approximations for
exchange and correlation.}
\label{density:plot}
\end{figure}

\begin{figure}[hbt]
\includegraphics[width=0.9\textwidth]{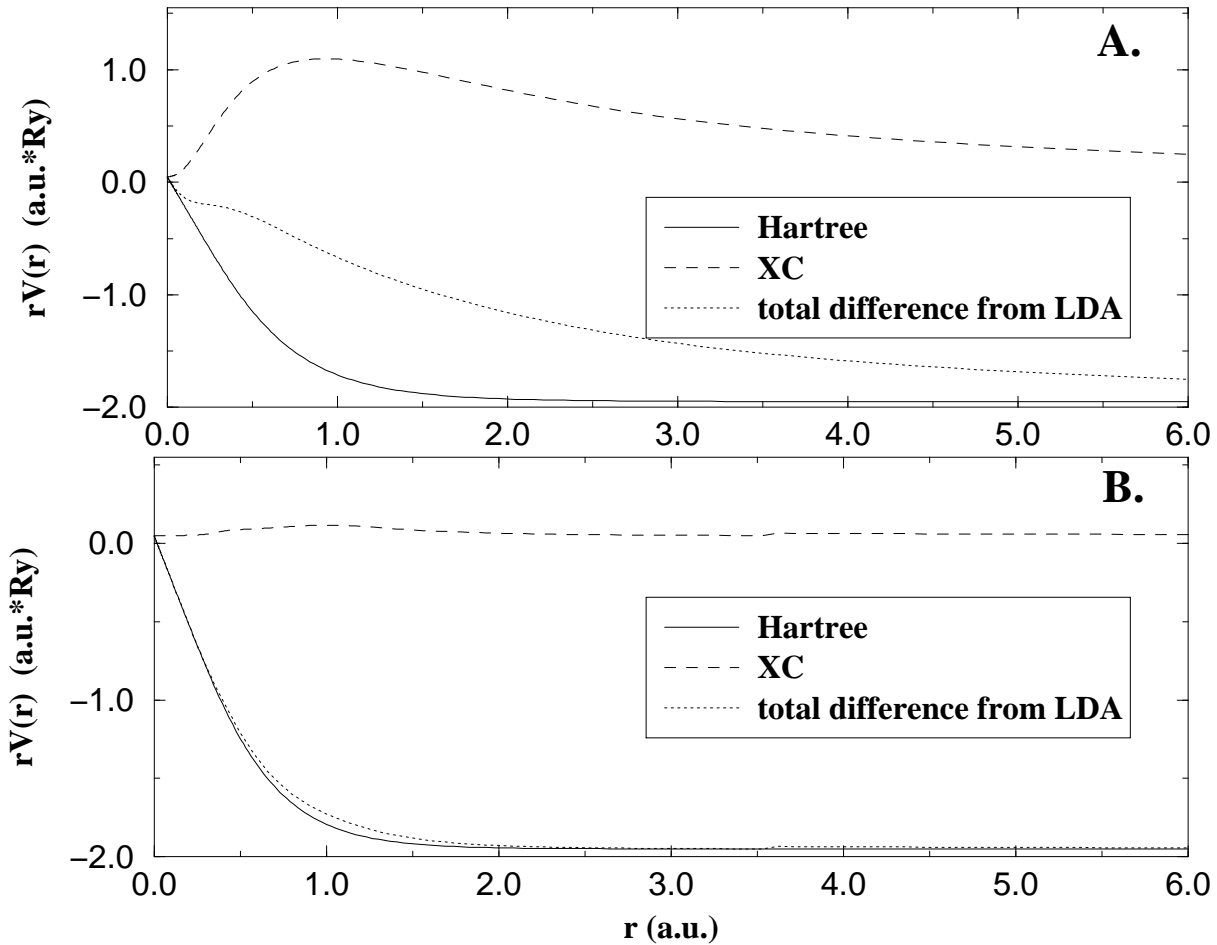}
\caption{
Fig.\ A shows the different contributions to the SIC potential in
the PZ-method. Fig.\ B shows the corresponding potentials for the
current implementation of SIC.}
\label{potential:plot}
\end{figure}

We have implemented this new method, and the method by PZ, to be able to
make a comparison between the two, in a full potential linear 
muffin-tin orbital (FP-LMTO) method\cite{wills}. 
Since Pr is a material with 2 localized f-electrons which has been studied
extensively by LDA\cite{min,melsen,delin}, 
PZ-SIC\cite{svane3,temmerman,szotek2,forstreuter}
and the orbital polarization method\cite{svane3} we
choose to test our theory for this element.
For simplicity Pr was put in an fcc structure (observed structure is the
dhcp structure), and  we used 2197 k-points 
in the whole 
Brillouin zone. The SI corrected functionals were implemented only for the 4f 
core electrons. In the implementation we have ignored
the non-orthogonality of the orbitals introduced by SIC 
since in atomic like 
problems it has been shown to be negligible\cite{perdew}.
The results for praseodymium 
are shown in the Table\ref{result:table} where the total energy
contributions are listed, and in 
Figures~\ref{density:plot} and \ref{potential:plot} where the electron
density and different contributions to the effective potential are
displayed.
\begin{table}
\label{result:table}
\begin{tabular}{c|ccc}
Quantity  & LDA & PZ-SIC\tablenotemark[1] & SIC\\ \hline
4$f$ KS-eigenvalue\tablenotemark[2] (Ry)
      & -0.054167  & -0.771376   & -1.537387 \\
Equilibrium volume\tablenotemark[3] ($R_{SWS}$ in a.u.)
      & 3.826        & 3.842     &    3.916        \\ \hline
Correction to energies  &      &   &          \\
Kinetic         (Ry) && -0.1239143 & -1.0536124    \\
Hartree          (Ry) && -1.8285461 & -2.0564759    \\
XC              (Ry) &&  1.7158233 & 0.522432025   \\ \hline
Total Energy    (Ry) & -18'471.575 781 & -18'471.825 455 & -18'473.480
380
\\
\end{tabular}
\tablenotetext[1]{SIC From Ref.\ \cite{perdew}.}
\tablenotetext[2]{With respect to the Fermi energy.}
\tablenotetext[3]{The experimental equilibrium volume is 3.818 a.u.}
\caption{Comparison between LDA, PZ-SIC and the current SIC method.
The energies come from the corresponding minima in volume. The energies
are the corrections to the (LDA) 4$f$-electron energies given by 
Eqn.(\ref{tote:correct}) for two 4f electrons.}
\end{table}
By inspecting the 4$f$-density (shown in Fig.\ref{density:plot}) 
one observes that
the orbitals become more contracted when SIC is used. Notable is also 
that the radial extent of the orbital in the new SIC method is largely 
reduced compared to the PZ-SIC method. We also note that compared to an atomic 
Hartree-Fock density the $f$-cloud is actually even more contracted.
From Fig.\ref{potential:plot}b one sees that for the current
SIC method the main contribution
to the potential entering the KS equation comes from the Hartree part, the
XC part provides a very small contribution. 
This can be compared
with the PZ-SIC potential shown in Fig.\ref{potential:plot}a, where 
{\it the same} Hartree part is used, but there is a big correction to the
XC potential. The reason why the correction to the 
XC potential in this implementation 
is so small is that the potential has the form
$V_{{\rm XC}}=V_{{\rm XC}}(\rho_{{\rm tot}}-\rho_i^\sigma)$ instead of the LDA 
$V_{{\rm XC}}=V_{{\rm XC}}(\rho_{{\rm tot}})$
and for large values of $\rho_{{\rm tot}}$ the difference between 
them, $\rho_{{\rm tot}}-\rho_i^\sigma ={\bar{\rho}} \approx \rho_{{\rm tot}}$, 
is very small. 
As a side remark we note that the electronic structure and 
density of states of the valence states is almost unchanged (no SIC is applied 
to the valent electrons), 
showing that the 
contraction of the 4\f cloud due to SIC only marginally influences the 
energy-band dispersion of the valence states. 

From the Table\ref{result:table} it is seen that the KS eigenvalue is 
significantly lower in the new SIC method compared to LDA and PZ-SIC.
The fact that the $f$-orbital localizes more gives rise to a larger 
screening for the valence electrons, which as a result become more 
extended and 
for this reason  a larger equilibrium volume is found. 
However, as is seen from the table the difference in equilibrium volumes 
between the different approximations is only a few percent and they agree 
rather well with experiment. The small  
deviation from experiment can be explained by a contribution to the 
cohesive energy from 
delocalized $f$-electrons as described previously\cite{lundin}.
The cohesive energy was previously calculated from LDA to be 5.09 
eV/atom\cite{delin}, a value which is 0.55 eV/atom higher than 
experiment. Our calculation lowers the cohesive energy 
with 0.29 eV/atom compared to the LDA calculation, and is hence in better 
agreement with experiment. The PZ calculation lowers the cohesive 
energy by only with 0.17 eV/atom compared to LDA.  

In the Table we also note that
the self-Hartree part is larger in our method of SIC than in the
PZ-method, this is due to the more contracted 4$f$-orbital.
In the PZ-SIC the total energy changes only by 0.25 Ry  
since the Hartree correction is balanced by a change in the XC energy.
In our implementation of SIC the Hartree energy is not balanced to the
same extent by the
XC energy, thus giving a change in total energy by 1.90 Ry.
We also compared the 4$f$-eigenvalue with an atomic, relativistic 
Hartree-Fock calculation on Pr$^{3+}$ with a 4f$^2$ configuration 
which is free from SI (but without correlation 
effects). This calculation gives a value of $\epsilon_{4f}$=-2.870 Ry, 
which may be compared to an 
atomic-like calculation (a band-calculation with lattice expanded) which gives
-2.3062 Ry for the new SIC method, -1.175 Ry for the PZ-SIC method and 
-0.444 Ry for a normal LDA calculation. Hence the current SIC method is
closer to a Hartree Fock description than LDA or PZ-SIC. This 
finding is consistent 
with the discussion above that Eqn.(\ref{effpot_new}) may be viewed as an 
approximation of the Hartree-Fock equation.

The present SIC method lowers the total energy compared to the method of
Perdew and Zunger. For this reason it may not be as suitable for
describing localization-delocalization transitions in the rare-earth
metals, since the PZ-SIC method reproduces the Mott transition in Ce
very well\cite{beiden,svane1}. However, since the PZ-SIC method does not
remove all self interaction it would seem that the good description
of this method in describing the localization-delocalization transitions
in the rare-earths is somewhat fortuitous. 

We also remark that 
as long as a Hartree term (which depends linearly on the electron
density) is introduced in the energy functional,
any local XC functional $\epsilon_{{\rm XC}}(\rho_i({\bf r}),0)$
which is a functional of the total density must 
also have a linear dependence on the
electron density, if SI is to be absent in the functional. 

By construction the LDA XC potential
cancels some of the unphysical self interaction, e.g. in hydrogen
95$\%$ of the self interaction\cite{jones} is cancelled. 
Application of the presented SIC method to the hydrogen-atom 
would yield an exact result (as would the PZ-SIC). 
Since the application of PZ-SIC normally improves the LDA result,
we believe that the currently described method of SIC will be 
successful when applied to 
systems like the 1-D Hubbard model and the bandgap problem in NiO, 
since it properly corrects for all self interaction. 
It may also be a good starting point for finding an exact
XC functional of localized or nearly localized electron
systems.

\section*{Conclusion}
In conclusion, we have presented a novel way to perform SIC-corrections in DFT 
calculations, and compared it to an older, not fully SI free, method. 
In its form the suggested correction is similar to the corrections
of the non-spherical 4f density proposed by Brooks {\it{et al}}. in
calculations of crystal field excitations\cite{brooks}.
The method presented is closer to the suggestion of Nesbet\cite{nesbet}
and differs from the method of e.\ g.\ Manoli {\it et al}\cite{manoli} 
whose approximation is more similar to that of Ref.\cite{perdew}. 
We have 
applied the new SIC method 
to praseodymium metal and showed that it results in stronger localization
of the f-shell,
thus correcting more for self interaction. Despite their large
differences in describing the 4f electrons, the calculated equilibrium
volumes of LDA, PZ-SIC and the presently proposed SIC method are all  
quite similar and agree with the experimental values. This is a result 
of that the valence states (that are not calculated using any SIC potential) 
are rather independent of the shape of the 4$f$-density and it is 
namely the valence states that are responsible for the chemical 
bonding. On the other hand, 
the cohesive energy is better described by the new SIC method, both 
compared to LDA and to PZ-SIC. 

\section*{ACKNOWLEDGMENT}
This project has been financed by
the Swedish Natural Science Research Council (NFR). 
We are also thankful to Sverker Edvardson
Mitth\"ogskolan, Sundsvall for providing results from the 
Hartree-Fock calculations.

\end{document}